\documentclass[aps,twocolumn,10pt]{revtex4} 
\usepackage{amsfonts} 
\usepackage{amsmath} 
\usepackage{amssymb} 
\usepackage{graphicx} 
\usepackage{epsfig} 

\begin{document} 
 
\title{Unconventional strongly interacting Bose-Einstein condensates in optical lattices} 
\author{ A. B. Kuklov } 
\affiliation{Department of Physics,CSI, CUNY - Staten Island, New York, NY 10314} 
 
\begin{abstract}
Feschbach resonances in a non-s-wave channel of two-component bosonic mixtures can induce atomic Bose Einstein condensates with a non-zero orbital momentum in the optical lattice, if one component is in the Mott insulator 
state and the other is not. 
Such non-s-wave condensates break the symmetry of the lattice and, in some cases, time-reversal symmetry. 
They can be revealed in specific absorption imaging patterns. 
\end{abstract} 

\date{\today}
\maketitle 
 
Unprecedented advances in trapping and manipulating ultracold atoms in optical
lattices (OLs) \cite{MI-SF,Bloch_noise} have created a growing wave of interest in strongly
correlated many-body phases and quantum phase transitions. The Mott insulator (MI) 
superfluid (SF) transition \cite{Fisher}  realized in one-component lattice bosons \cite{MI-SF}
has inspired predictions of the variety of novel phases 
\cite{novel_phases} 
in quantum mixtures. These phases are all described by
  the Hubbard-type Hamiltonian \cite{Fisher} in which the lowest single-particle band only
is considered in the limit when a typical one-particle band gap exceeds considerably 
the on site interaction energy. 
Recently the opposite limit has been reached \cite{ETH_1} in the s-wave channel. 
As pointed out by Diener and Ho in ref.\cite{HO_1}, the resulting effect of the
interaction induced band mixing \cite{ETH_1} can potentially be a source of yet novel
quantum phases and phase transitions. The effects of macroscopic bosonic population
of higher Bloch bands under strongly non-equilibrium conditions have been analyzed
in detail in ref.\cite{Girvin}. At commensurate filling the insulator-superfluid transition
can also be observed in the higher bands \cite{Girvin}. 

In this Letter, it is shown that Feshbach resonance at 
some $L\neq 0$ angular momentum
can induce {\it atomic} Bose Einstein condensate (BEC) which breaks crystal symmetries
by means of essentially equilibrium phase transition.
 Such non-s-wave BEC must be contrasted, on one hand,
 with the conventional (s-wave) BEC carrying no angular momentum in its ground state 
 and, on the other hand, with
the {\it paired}  states in the fermionic systems: $^3$He and
unconventional superconductors \cite{non-s-wave,Sigrist_2}. 
The p-wave resonances have been observed in one component
fermions \cite{p-wave,p-split}. Higher-$L$ resonances have been detected
in bosonic $^{87}$Rb \cite{d-wave} and $^{133}$Cs \cite{high-wave} systems as well. In the latter case spin-orbit
coupling creates interference between different $L$ channels. Here, however,
the case of isolated resonance with $L>0$ is considered for two {\it distinguishable} bosonic atoms
in OLs in order to discuss novel phases and phase transitions in the simplest formulation.

The non-s-wave BECs are characterized by zeros and degeneracy. 
This apparently contradicts to  Feynman's argument
that the bosonic many-body ground state must have no zeros \cite{Feynman}. 
In the case $L\neq 0$, however, the system is not in its 
absolute ground state due to the closed-channel molecular component.  Thus, a such  BEC must be viewed in the 
context of metastable phases generic for atomic traps and OLs, with the stability insured
by large  difference between intra-molecular and atomic center of mass energies.  
Accordingly, the system can be tuned to exhibit quantum phase transitions between
the traditional and the orbital BECs.

For two {\it distinguishable and non-convertible} bosons, e.g., A and B, in OLs,
each site can be approximated by superimposed isotropic oscillator potentials 
with frequencies $\omega_a,\, \omega_b$ and masses
$m_a,\, m_b$ for the sorts A and B, respectively. 
It is reasonable to choose $m_a=m_b=m$ and
$\omega_a=\omega_b=\omega_0$ in order to allow decoupling bewteen center of mass
and relative motions.
The one-particle states are classified according to the radial number and
to the value of the angular momentum. The lowest states 
\begin{equation} 
1s,\,\, (1p), \,\, (1d,2s)
\label{Land} 
\end{equation} 
 are shown in order of increasing energy from left to right.
The states $1s,\, 2s$ carry no agular momentum $L$.
The state $1p$ ($L=1$) is triple degenerate and transforms as a vector
\begin{equation} 
\psi_{1p} \equiv \psi_i \,\, \sim \,\, x_i,
\label{1p} 
\end{equation}
where $x_i=x,\, y,\, z$ stand for Cartesian coordinates.
The state $1d$ with $L=2$ is fivefold
degenerate and transforms as a tensor: 
$\psi_{1d} \equiv \psi_{ij} \,\, \sim \,\, x_ix_j - \frac{{\bf x}^2}{3}\delta_{ij}$.

Let's consider a situation of  Feschbach resonance
in the $L$-wave channel between the  species. The corresponding closed channel molecular state 
$\Phi_\alpha({\bf r})$ is characterized by some degeneracy index $\alpha$ (for $L=1$, $\alpha =x,y,z$  \cite{note_1}) and $\bf r$ stands for the relative
coordinate of the A and B atoms.
 The analysis for one site 
closely follows the approach \cite{HO_1}. Its validity for the lattice is determined
by smallness of a typical intersite tunneling amplitude $t$ in comparison with the on site excitation energy $\omega_0$. 
The size $r_\phi$ of the molecule is assumed
to be much less than the oscillator length $r_0=1/\sqrt{m\omega_0}$ (units $\hbar=1$ are employed here and below) \cite{Chevy}. 
Introducing the creation-annihilation operators for
the molecule $c^\dagger_\alpha,\, c_\alpha $ and for the A and B bosons  $a^\dagger_m,\,
a_m$ and $b^\dagger_n,\, b_n$, respectively, with $m,n$ referring to the oscillator levels,
the onsite Hamiltonian can be written in the form
\begin{eqnarray} 
H&=&  \sum_\alpha \bar{\nu}c^\dagger_\alpha c_\alpha + \sum_{m}[\varepsilon_{am}a^\dagger_ma_m + \varepsilon_{bm}b^\dagger_mb_m]
\nonumber \\
&-& \sum_{m,n} \gamma_{\alpha,mn}(c^\dagger_\alpha a_mb_n + H.c.),
\label{H} 
\end{eqnarray}
 with $\bar{\nu}$ standing for the bare molecular energy and
$\varepsilon_{am},\, \varepsilon_{bm}$ labeling the oscillator energy levels. Here particle interactions
beyond the resonant one are not shown explicitly. The resonance-interaction matrix elements $\gamma_{\alpha,mn}$  
are determined by the strength of the open-closed channels
coupling and by the symmetry: for given $\alpha$, it selects such states $m,\, n$ so
that the center of mass of the pair resides 
in the $1s$ oscillator state $\sim \exp(-m \omega_0 {\bf R}^2)$ and their
relative motion is characterized by oscillator wavefunction $\psi_n({\bf r})$ with given $L$. For example,
for $L=1$ the lowest energy term is the $1p$ state $\psi_{n=\alpha}({\bf r})\sim \exp(-m \omega_0 {\bf r}^2/4)({\bf r})_\alpha,\,\, \alpha=x,y,z$.
Here (for equal masses), $\bf R=({\bf r}_a + {\bf r}_b)/2,\,\, {\bf r}={\bf r}_a -{\bf r}_b$  and
${\bf r}_a$ and ${\bf r}_b$ refer to the coordinates of the A and B particles, respectively. 

For large enough $\gamma_{\alpha,mn}$ it is natural to expect that, upon proper tuning
of $\bar{\nu}$, it is possible to make the lowest energy $E_L$ of the AB-pair with relative $L$ to lie {\it below} 
the energy $E_0=\varepsilon_{a0} + \varepsilon_{b0}=3\omega_0$ 
of the $1s$ state for both atoms  $|L=0\rangle=a^\dagger_0b^\dagger_0|0\rangle$ (in which that they
do not participate in the resonance).
This is possible
because no states with $L=0$ are simply present in the expansion given by $\gamma_{\alpha,mn}$.
The ground state should be looked for in the form \cite{HO_1}:
\begin{equation} 
|L,\alpha\rangle=B^\dagger_{L\alpha}|0\rangle,\quad B^\dagger_{L\alpha}= \beta c^\dagger + \sum_{mn}\eta_{mn}a^\dagger_mb^\dagger_n ,
\label{gen} 
\end{equation} 
with $\beta^2 + \sum_{mn}\eta_{mn}^2=1$, and the summation running over the
oscillator states corresponding to the $\alpha$ orbital of the
chosen $L$. 
Then, $E_L$ must be compared with $E_0$. 
The full procedure of finding $\beta, \eta_{mn}, E_L$
requires proper renormalization of the detuning $\bar{\nu}$ \cite{HO_1}.
Some complication also comes  from 
the sensitivity of the eigenenergies to microscopic details of $\Phi_\alpha({\bf r})$.
One can, however, choose the variational approach in order to obtain the {\it upper bound}
estimate for $E_L$ in terms of the matrix elements
$\gamma_{\alpha,mn}$ admixing the lowest oscillator levels only.
Hence, if the condition $E_L<E_0$ holds for the upper bound of $E_L$, it
will be satisfied within the exact scheme \cite{HO_1} for sure.
This also helps establishing the symmetry and topological properties of the phase because
these do not depend on the number of states involved in the expansion (\ref{gen}). 

In the case $L=1$, the lowest relevant oscillator states are $1s$ and $1p$ from
the nomenclature (\ref{Land}). 
Then, the two-body wavefunction in the open channel
$\Psi_{0\alpha} ({\bf r}_a,{\bf r}_b)\sim \exp(-m \omega_0 {\bf R}^2)
 \exp(-m \omega_0 {\bf r}^2/4)({\bf r})_\alpha$ can be expanded as
$\sim \exp(-m \omega_0 {\bf r}_a^2/2)({\bf r}_a)_\alpha
 \exp(-m \omega_0 {\bf r}_b^2/2) - \exp(-m \omega_0 {\bf r}_b^2/2)({\bf r}_b)_\alpha
 \exp(-m \omega_0 {\bf r}_a^2/2)$. This  translates into truncating the expansion
(\ref{gen}) by the lowest terms only: 
$|L=1,\alpha\rangle=
[\beta c^\dagger + \eta(a^\dagger_0b^\dagger_\alpha-b^\dagger_0a^\dagger_\alpha)] |0\rangle$, with
$\beta^2 + 2\eta^2=1$. Then, $E_L=\langle L=1,\alpha|H|\alpha, L=1\rangle$ selects
only the terms with
$\gamma_{\alpha,0\alpha}=- \gamma_{\alpha,\alpha 0}=\gamma$ in eq.(\ref{H}).
The resulting {\it upper bound} estimate for the ground
state energy is $E_L=(\bar{\nu} + 4\omega_0)/2 -\sqrt{(\bar{\nu} - 4\omega_0)^2/4 +\gamma^2}$.
If $\bar{\nu}$ is tuned to $\approx 4\omega_0$, then $E_L\approx 4\omega_0 -|\gamma|$ and
the requirement $E_L < E_0$ 
translates into
\begin{equation}
|\gamma| \geq \omega_0,
\label{gamma}
\end{equation}
so that $|L=1,\alpha\rangle$ becomes the ground state.  Tuning
$\bar \nu$ (or its renormalized value  \cite{HO_1}) to higher values will eventually
raise the energy $E_L$ above $E_0$. In contrast, if $|\gamma| < \omega_0$, no effect is most likely to be
observed for any ${\bar \nu}$.

The value of 
$\gamma$ is determined by exchange interaction $V_{ex}$ on molecular distances further
reduced by small overlap between the closed and open channels.
Using $V_{ex}\sim$ 1eV on atomic distances $a_0\sim 10^{-8}$cm
and taking $r_\phi \sim 1$nm \cite{Chevy} and $r_0\sim 10^{-4}$cm,
the estimate for arbitrary $L$ gives $\gamma \sim V_{ex} (a_0/r_\phi) (r_\phi/r_0)^{3/2 + L}$,
 where 
it was taken into account that the wave function of the relative motion is $\sim r^L$ at the origin.
For the p-wave case ($L=1$) one finds
$\gamma \approx 30 \mu$K$\approx 3$MHz.
This shows plenty of room for satisfying the condition (\ref{gamma}) for typical OL frequencies
in the range of few kHz.  Clearly, higher resonances with $L=2,3...$ will be characterized by smaller $\gamma$.
In terms of the
typical exchange matrix element $\tilde{V}_{ex}=V_{ex}r_\phi/a_0$, the typical molecular kinetic energy $\epsilon_0 \approx
1/(mr^2_\phi)$ and $r_\phi/r_0$ the estimate as a function of $L$ becomes 
\begin{equation}
\gamma /\omega_0 \approx
(\tilde{V}_{ex}/\epsilon_0)(r_\phi/r_0)^{L-1/2}.
\label{gamma_2}
\end{equation}
 For chosen parameters, this gives $\gamma /\omega_0 \approx
10^5(r_\phi/r_0)^{L-1/2}$, and the condition (\ref{gamma}) 
can be satisfied by $L=1,2$. Choosing to work with smaller $r_0$, about $r_0\approx 10^{-5}$cm, will allow
realization of the ground state with $L$ as high as $L=3$. For higher $L$, however, it becomes important
that the one particle band gap is $L\omega_0$, and the condition (\ref{gamma}) must, actually, 
be $\gamma \geq L\omega_0$.

{\it Non-s-wave Bose-Einstein condensate of the A+B mixture}.---Starting from the
deep MI "vacuum" of A-bosons 
$|MI\rangle =\prod^N_i a^\dagger_0(i)|0\rangle$ in the lattice of $N$ sites, one can
add a small number $N_B \ll N$ of B-bosons in the regime of the $L$-wave A+B
resonance. This will result in formation of the 
onsite states (\ref{gen}). In the limit of weak tunneling
such resonant complexes will delocalize over the lattice without changing their shape. 

The state (\ref{gen}) is characterized by strong entanglement, and
it is important to verify that this is not in a conflict 
with the formation of the off-diagonal long range order (ODLRO). The operator 
$B^\dagger_{L\alpha}(i)a_0(i)$
creates one excitation on a site $i$ from $|MI\rangle$.
 Here and below the dependence
on the site index $i$ is shown explicitly.
For low density of the excitations, it is
enough to verify that one excitation forming a band wave
\begin{equation} 
{\tilde B}^\dagger_{L\alpha}=\sum_i \frac{1}{\sqrt{N}}B^\dagger_{L\alpha}(i)a_0(i)
\label{B_band}
\end{equation}
contributes constructively to the ODLRO density matrix
$\rho^{(b)}_{mn}(i,j)=\langle MI| {\tilde B}_{L\alpha} b^\dagger_m(i) b_n(j) {\tilde B}^\dagger_{L\alpha}|MI\rangle$
in the limit $|{\bf x}_i -{\bf x}_j| \to \infty$. Direct calculation 
gives $\rho^{(b)}_{mn}(i,j)= \eta^*_{0n}\eta_{0m}/N$.
For $1\ll N_B \ll N$:
\begin{equation} 
\rho^{(b)}_{mn}(i,j)= \eta^*_{0n}\eta_{0m} n_B ,
\label{rho}
\end{equation}
where  $n_B= N_B/N$, with the index $n$ referring to the 
onsite oscillator states with the selected (by the resonance) angular momentum $L$ of a B-particle.
It is worth noting quite significant quantum depletion of the condensate density
even in the limit $n_B \ll 1 $ due to $|\eta_{0n}|<1$. 
Within the variational approach discussed above, $\eta_{0n}$ is to be replaced by
$\eta$, and the indices $m,n$ must be set equal to a given $\alpha$-orbital of the, e.g.,
$1p$ oscillator state \cite{note5}.
It is important to note that no other channel contains the ODLRO, that is,
$\langle b^\dagger_0(i) b_0(j)\rangle =0,\, 
 \langle a^\dagger_m(i) a_n(j)\rangle=0$ and $\langle c^\dagger_\alpha(i) c_{\alpha'}(j)\rangle=0$ for ${\bf x}_i \neq {\bf x}_j$.

{\it Macroscopic properties of the p-wave BEC}.---Any $L>0$-wave BEC
have {\it zero } population of the zero-momentum state
even in an infinite and uniform system.
This follows from vanishing of the space integral of any non-s orbital function. 
In other words, $\int d{\bf x} W_\alpha ({\bf x}- {\bf x}_i)=0$
where $W_\alpha ({\bf x}- {\bf x}_i)$ stands for a non-s-wave  Wannier function 
localized at the coordinate ${\bf x}_i$ of the $i$th site.
This leads to quite unusual macroscopic properties \cite{Girvin}. 

Let's consider, first, two neighboring sites, say, $i$ and $j$. Then, the tunneling
part of the Hamiltonian can be constructed following the standard procedure
\cite{Zoller_1}. The symmetry condition (\ref{1p}) immediately implies
that no tunneling takes place between states with different $\alpha=x,y,z$. Furthermore,
since $W_\alpha({\bf x})$ is odd for the $p$-wave, the tunneling amplitude ~$t^{(\alpha)}_{ij}$
is {\it negative} along the orientation
of the $\alpha$-orbital and positive if perpendicular. The reason for that is that the 
combination $W_\alpha({\bf x} - {\bf x}_i) + W_\alpha({\bf x} - {\bf x}_j)$ 
for $\alpha$ being along the direction $\sim {\bf x}_i - {\bf x}_j$
has one extra zero in comparison with the one
 $W_\alpha({\bf x} - {\bf x}_i) - W_\alpha({\bf x} - {\bf x}_j)$.
Thus,
the lowest energy of the secular equation corresponds to the
antisymmetric combination. 
In general, the sign of ~$t^{(\alpha)}_{ij}$  depends on mutual orientation of the
vector ${\bf u}=({\bf x}_i - {\bf x}_j)/|{\bf x}_i - {\bf x}_j|$ and of the orbital. If
 $u_x=1,\, u_y=0,\, u_z=0$ and the orbital assignment is taken
according to the $x$-axis being along ${\bf u}$, then, there are two non-zero
values in the set $t^{(\alpha)}_{ij}$. These are $t_{||}=t^{(x)}_{ij} < 0 $ and
$t_\perp =t^{(y)}_{ij}=t^{(z)}_{ij} >0 $.

The lattice Hamiltonian formulated
in terms of the lowest $L=1$-band operators $b^\dagger_\alpha(i),\,b_\alpha(i)$,becomes
\begin{equation} 
H_B= - \sum_{\alpha,<ij>} t^{(\alpha)}_{ij}b^\dagger_\alpha(i)b_\alpha(j) - 
\sum_{i, \alpha \beta}{\bf \Omega}{\bf L}_{\alpha \beta}b^\dagger_\alpha(i)b_\beta(i)
\label{B-tun}
\end{equation}
in the hard-core approximation $b_\alpha(i)b_\beta(i)=0$. 
The last term in eq.(\ref{B-tun}) describes the effect of external
rotation of the lattice at some angular frequency $\bf \Omega$, with ${\bf L}_{\alpha \beta}$ standing
for the angular momentum operator $\hat{\bf L} $ built on the on site Wannier functions:
 ${\bf L}_{\alpha \beta}=\int d{\bf x} W^*_\alpha({\bf x})\hat{\bf L}W_\beta({\bf x})$. 
Thus, in addition to the possibility of introducing vortices, external rotation induces mixing between 
the components. The mixing can also be induced by the exchange onsite interactions \cite{Girvin}.

In general, properties of the non-s-wave BEC depend strongly on the lattice
symmetry and its structural defects and boundaries. In particular, the translational
lattice symmetry is broken. For the p-wave BEC,
 the phase $\varphi_r$ in $\langle b_{\alpha}(r)\rangle \sim {\rm e}^{i\varphi_r}$
must oscillate as $\varphi_r= \pi (1+(-1)^r)/2$ along the direction of the orbital
in order to lower the tunneling energy. Such a tendency to form
spatial superstructure of the phase can lead 
to {\it spontaneous} ground state currents in, e.g., the triangular lattice
because of the sign frustration along the elementary plaquet.

In the tetragonal lattice the condensation proceeds into either $p=z$ 
or the $p=x,y$ states. 
If the $z$-axis lattice constant is the shortest one, 
the tunneling energy will be gained by forming the one-component
$z$-BEC. In the opposite case, 
the two-component $xy$-BEC occurs with two options: (i) The order parameter (OP) is
 real of the type $ x \pm y$; (ii) the OP is complex as $x\pm iy$ and, therefore, it breaks the time-reversal
symmetry. What option is actually realized depends on the details of the interaction.
On the mean field level, in the $xy$-BEC two components $\psi_x$ and $\psi_y$
must be considered \cite{Sigrist_2}. The corresponding Landau expansion obeys
the lattice symmetry. Thus, the quadratic  $|\psi_{x,y}|^2 $
and the gradient terms do not distinguish between the options. Among the
quartic terms, there is one $H_t= \int d^3x g_t (\psi^{*2}_x \psi^2_y + c.c.)$
which determines the phase locking between the components similarly to the situation
in the d-wave superconductor \cite{Sigrist_1}. If $g_t <0$, the option (i) takes place.Then,
the phases $\varphi_x$ and $\varphi_y$ are equal to each other and the condensation
proceeds into the state $\psi_x \pm \psi_y$ with $|\psi_y|=|\psi_x|=\sqrt{\rho_p}$.
Accordingly, the line of zeros will be arranged along the plane $x=\mp y$. 
If, however, $g_t >0$, the energy can be lowered if $\varphi_x - \varphi_y= \pm \pi/2$, so that
there are no zeros in $\psi_\pm= \psi_x \pm i\psi_y$.
Then, similarly to the case of the d-wave superconductor \cite{Sigrist_1},
the ground state can have spontaneous fractional vortices initiated close to the
OLs boundaries and structural defects. 

The Josephson effect can also feature
spontaneous currents \cite{Sigrist_1} even though the time-reversal is not broken. 
If the $p$-orbital BEC is surrounded by the s-wave
BEC of the same particles, the tunneling current $J_c(\theta)$  between two BECs will strongly depend 
on the  angle $\theta$ between the orientations of the orbital ${\bf u}$ and of the boundary
${\bf n}^{(s)}$. 
  If ${\bf u}\cdot {\bf n}^{(s)}=0$, no tunneling will take
place because the current from the positive part of the lobe of the $p$-orbital
will be exactly compensated by the current from the negative part.
In general, $J_c(\theta)\sim \cos \theta$.
Thus, there will be a phase jump of $\pi$ for a contour traversing both SFs
so that $\theta$ changes by $180^o$ along it.
This implies that a spontaneous vortex,  carrying
1/2 of the standard circulation, can form if the system size $R$ is macroscopic
in comparison with the healing length and any length associated
with the Josephson coupling. 

{\it $L$-wave Mott insulator}.---At $N_B=N$ and for strong s-wave intraspecies repulsion the system
becomes MI with broken lattice (including time-reversal) symmetries (compare with
the single boson case \cite{Girvin}).
 Obviously,
tuning away from the $L$-wave resonance
will result in a phase transition into the two-component MI
with no broken symmetries. The nature of the $L$-wave MI as well as of
the phase transition requires separate analysis.

{\it Detection}.---Non-s-wave condensates should produce specific absorption
imaging patterns due to condensation into the ${\bf k}\neq 0$ state \cite{Girvin}.
For example, considering the options (i) and (ii), the $\psi_{a=1,2}$ operator
becomes
\begin{equation}
\psi_a ({\bf x})=\sum_i W_a({\bf x} - {\bf x}_i) b_a(i),\,\, a=1,2.
\label{expand}
\end{equation} 
The Wannier function can be taken as 
$W_1({\bf x})\approx a_w^{-5/2}\,e^{-{\bf x}^2/2\,a_w^2} (x \pm y)$,
with $a_w \approx r_0$,
in the case of the time-invariant state (i) and as
$W_2({\bf x})\approx a_w^{-5/2}\,e^{-{\bf x}^2/2\,a_w^2} (x \pm iy)$
 in the case (ii) of the broken time-reversal symmetry.
After atoms are released from the lattice and some free expansion, the cloud density $n_B({\bf x},t)$ of
the B-atoms
in the far zone becomes \cite{image}: $
n_B({\bf x},t) \sim |\tilde{W}_a({\bf q})|^2 \sum_{ij}\exp(i{\bf q}({\bf x}_i - {\bf x}_j))\langle b^\dagger_a(i)b_a(j)\rangle$,
where $\tilde{W}_a({\bf q})$ stands for the Fourier transform 
of $W_a({\bf x})$ with the momentum ${\bf q}= m{\bf x}/t$. 
These transforms ~$
\tilde{W}_1({\bf q})\approx t^{-5/2} (x\pm y) $ 
and
~$\tilde{W}_2({\bf q})\approx t^{-5/2} (x\pm iy) $ 
should be
contrasted with the one for the s-wave BEC, $\sim t^{-3/2}$ 
The expressions $\tilde{W}_{1,2}({\bf q})$ feature modulation of the standard Bragg peaks \cite{MI-SF} 
by the factors $(x\pm y)^2/t^2$
and $(x^2 +y ^2)/t^2$ for the options (i) and (ii), respectively. In the first case, the peaks along the line (in the columnar image) $x=\mp y$ will be suppressed. In the second, the central Bragg spot only will be strongly suppressed. 
In general, the imaging pattern reflects the symmetry of the orbital and of the lattice.
[Similar expression can be easily obtained for the p-wave-BEC with the OP of the $z$-type.]
The corresponding patterns are recognizable by the lines of zeros or by the suppression
of the central Bragg peak. Higher $L$ condensates will feature more complex patterns.
It is worth noting that rotation of the lattice can produce switching between the patterns, e.g.,
from (i) to (ii) because finite $\bf \Omega$ in eq.(\ref{B-tun}) promotes ferromagnetic ordering of the on site angular momenta -- the state (ii): $\psi_\pm \sim x\pm iy$, with the sign chosen
by the direction of $\bf \Omega$.

In conclusion, Feshbach resonance in a non-zero angular momentum channel 
favors formation of the non-s-wave BEC characterized by space-time symmetries lower than
those of the OLs. The ground state contains lines of zeros and spontaneous currents induced
by boundaries and structural defects. The absorption images exhibit additional strong
modulational patterns reflecting the non-trivial symmetry of the ground state.
Imposing non-s-wave Feschbach resonance on a lattice with double occupation in MI
regime can induce insulating phases with broken lattice
symmetry. Varying the detuning of non-s-wave resonance can lead to 
various quantum phase transitions between the phases: S-wave BEC, non-S-wave BEC,  
conventional MI and orbital MI (with broken lattice symmetries).

Similar conclusions about properties of the p-orbital
BEC have been obtained in Refs.\cite{Liu}, where the case of a boson-fermion
mixture in OLs was analyzed.

This work was  supported by NSF, Grant PHY-0426814, and by the PSC, Grant CUNY-66556-0036. 
The author acknowledges hospitality of the Aspen Center for Physics during the Summer Workshop on ultra-cold atoms
when this work was initiated.


\begin{references} 
\bibitem{MI-SF} M. Greiner, et al. 
Nature {\bf 415}, 39(2002).

\bibitem{Bloch_noise}
S.F\"olling, et al., 
Nature {\bf 434}, 481 (2005) 
 
\bibitem{Fisher}
M. P. A. Fisher, et al. 
Phys. Rev. B {\bf 40}, 546 (1989)

\bibitem{novel_phases} E. Demler and F. Zhou 
Phys. Rev. Lett. {\bf 88}, 163001(2002); 
A.B. Kuklov and B.V. Svistunov, Phys. Rev. Lett. {\bf 90}, 100401(2003);
L.-M. Duan, E. Demler, and M.D. Lukin, Phys. Rev. Lett. {\bf 91}, 090402(2003). 
M. Lewenstein, et al. 
Phys. Rev. Lett. {\bf 92}, 050401 (2004).
 
\bibitem{ETH_1}
M. K\"ohl, et al. 
Phys. Rev. Lett. {\bf 94}, 080403 (2005)

\bibitem{HO_1}
R. B. Diener, T.-L. Ho, Phys. Rev. Lett. {\bf 96}, 010402 (2006).

\bibitem{Girvin}
A. Isacsson and S. M. Girvin, Phys. Rev. A {\bf 72}, 053604 (2005).

\bibitem{non-s-wave}
E. M. Lifshitz, L. P. Pitaevskii,  Ch.V.54, {\it Statistical Physics. II}, Pergamon (1980);
P. W. Anderson and P. Morel,
Phys. Rev. Lett. {\bf 5}, 136 (1960); ibid, {\bf 5}, 282 (1960);
Phys. Rev. {\bf 123}, 1911 (1961).



\bibitem{Sigrist_2}
M. Sigrist and K. Ueda, Rev. Mod. Phys. {\bf 63}, 239 (1991).

\bibitem{p-wave}
C. A. Regal, et al., 
Phys. Rev. Lett. {\bf 90}, 053201 (2003).

\bibitem{p-split}
C. Ticknor, et. al., 
Phys. Rev. A {\bf 69}, 042712 (2004). 

\bibitem{d-wave}
T. Volz, et al., 
Phys. Rev. A {\bf 72}, 010704(R) (2005).

\bibitem{high-wave}
C. Chin, et al., 
Phys. Rev. Lett. {\bf 85}, 2717 (2000); 
P. J. Leo, C. J. Williams, and P. S. Julienne, ibid,
p. 2721.



\bibitem{Feynman}
R.P. Feynman, {\it Statistical Mechanics. A set of lectures}, Frontiers
of Physics, Ch. 11.3. Ed: D. Pines. Reading-Tokyo, 1976.

\bibitem{note_1}
This degeneracy can be removed 
by magnetic dipole \cite{p-split} or spin-orbit interactions. While introducing some
complexity into the analysis, such circumstance does not change the main conclusions. 

\bibitem{Chevy}
F. Chevy, et al., Phys.Rev. A {\bf 71}, 062710 (2005).

\bibitem{note5}
The presence of higher excited states with given $L$ and $\alpha$ in $\rho^{(b)}_{mn}(i,j)$ 
is essentially the same effect discussed in Ref.\cite{HO_1} for the s-wave fermionic resonance . 
Since all bands with given $L$ have the same symmetry, 
it is enough to consider the lowest one only. 

\bibitem{Zoller_1} 
D. Jaksch, et al., 
Phys. Rev. Lett. 81, 3108-3111 (1998).



 
\bibitem{Sigrist_1}
M. Sigrist, D. B. Bailey, and R. B. Laughlin
Phys. Rev. Lett. {\bf 74}, 3249 (1995).

\bibitem{image}
V.A. Kashurnikov, N.V. Prokof'ev, and B.V. Svistunov, Phys. Rev. A {\bf 66}, 031601(R) (2002).

\bibitem{Liu}
W. V. Liu, C. Wu, Phys. Rev.A {\bf 74}, 013607 (2006);
 C. Wu, et al., cond-mat/0606743. 
 

\end{references}
\end{document}